\newcommand{\bear}{\begin{array}}  \newcommand{\eear}{\end{array}}
\newcommand{\bea}{\begin{eqnarray}}  \newcommand{\eea}{\end{eqnarray}}
\newcommand{\beq}{\begin{equation}}  \newcommand{\eeq}{\end{equation}}
\newcommand{\bef}{\begin{figure}}  \newcommand{\eef}{\end{figure}}
\newcommand{\bec}{\begin{center}}  \newcommand{\eec}{\end{center}}

\newcommand{\diracslash}[1]{#1\llap{/\kern2pt}}

\def\bearr{\begin{eqnarray}}
\def\eearr{\end{eqnarray}}

\documentclass[preprint,onecolumns,secnumarabic,floats,amssymb,
nobibnotes,nofootinbib,aps,prc,showpacs,tightenlines]{revtex4}
\usepackage{shrthnds}
\usepackage{amssymb}
\usepackage{graphicx}
\usepackage{amsmath}
\usepackage{latexsym}
\usepackage{epsfig}
\usepackage{subfigure}
\usepackage[dvipsnames]{xcolor}
\begin{document}

\title{ On existence of a possible Lorentz invariant modified gravity in Weitzenb\"{o}ck spacetime}
\author{
Davood Momeni$^{}$\footnote{email: d.momeni@yahoo.com},
Ratbay Myrzakulov$^{}$\footnote{email: rmyrzakulov@gmail.com}}
\affiliation{Eurasian International Center for Theoretical Physics and Department of
General \& Theoretical Physics, Eurasian National University, Astana 010008, Kazakhstan}
\date{\today}
%\date{}

%%
\vspace{1.5 cm}
\begin{abstract}
\vspace{1.5 cm}
Modified gravity which was constructed by torsion scalar $T$, namely $f(T)$ doesn't respect Lorentz symmetry. As an attempt to make a new torsion based modified gravity with Lorentz invarianve, recently $f(T,\mathcal{B})$ introduced where $B=2\nabla_{\mu}T^{\mu}$ \citep{Bahamonde:2015zma}. We would argue, even when all is constructed and done in a self-consistent form, if you handle them properly,we observe that there is no Lorentz invariant  teleparallel equivalent of $f(R)$ gravity. All we found is that the $f(R)$ gravity in which $R$ must be computed in Weitzenb\"{o}ck spacetime, using Weitzenb\"{o}ck's connection, nor Levi-Civita connections is the only possible Lorentz invariant type of modified gravity. Consequently, $f(T)$ gravity can not obey Lorentz symmetry not only in its orthodoxica form  but even in this new framework $f(T,\mathcal{B})$.
\end{abstract}

\maketitle
%%%%%%%%%%%%%%%%%%%%%%%%%%%%
\section{Introduction}
%%%%%%%%%%%%%%%%%%%%%%%%%%%%%%%
At this time we have several unsolved problems in cosmology and it is clearly demonstrated that the classical relativistic cosmology which 
 is based on the purely "classical" gauge theory is not able to resolve contemporary problems of the modern physics. One of the most important problems to be addressed adequately is to explain the origion and essence of an "unusual" currently observed acceleration expansion of the whole Universe. This type of acceleration is proposed based on 
different obervational data 
 as type Ia supernovae , cosmic microwave background (CMB) , large scale structure , baryon acoustic oscillations , and weak lensing  indicate \cite{data}. Modification(s) of the Einstein-Hilbert gravitational action is an interesting  approach to resolve the problem of dark energy as well as dark matter. This approach is known as modified gravity in which not only action but also dynamics will be changed dranatically(
for a review of modified gravities see \cite{review}). Several possibilities exist for modified gravity. The first model of modified gravity proposed by Buchdahl as an attempt to generalize Einstein gravity to an arbitrary function of 
 the Ricci scalar "R" , namely is called as $f (R)$, \cite{Buchdahl:1983zz}:
\begin{eqnarray}
S=\int{f(R) \sqrt{-g}d^4x} \label{f(R)}.
\end{eqnarray}
This type of modified gravity has two important features: one is handy to work from the mathematical point of view and secondly is this theory has 
Lorentz invariance. This model of gravity   is motivated recently in diverse forms by several authors 
\cite{f(R)}). The old proposal of $f(R)$ gravity mimics gravity as a minimal  modification. The reason is that $R$ in this theory is constracted from the metric and its first and second derivatives. So, theory is second order with respect to the metric. But from the geometrical point of view , there are several higher order curvature corrections which are inspired from string theory and high energy physics. The first higher order term is 
the Gauss-Bonnet (GB) $G=R^2-4R_{\mu\nu}R^{\mu\nu}+ R_{\mu\nu \lambda\sigma}R^{\mu\nu\lambda\sigma}$. This type of modified gravity is developed as parallel as $f(R)$ gravity . We know that so far,  we have several intereating models of modified gravity which are constructed from both $R,G$
quantities\cite{GB}. As a fact, it was demonstrated that the dynamics of cosmological backgrounds will be changed drastically by such higher order curvature corrections
\cite{Nojiri:2005jg,Cognola:2006eg}.
\par
From the historical point of the view, the metric of a generic spacetime can define two types of the connection terms (the terms which are constructed from the first order derivatives of metric $\partial_{\mu}g^{\alpha\beta}$. In general relativity (GR) we only use the symmetric part of it, the one called Levi-Civita (LC) connection $\Gamma^{\alpha}_{\mu\nu}$. The other one which constructed by the assymmetric part, is called Weitzenb\"{o}ck connection is also provided a consistent formalism to describe gravity 
s gauge theory. This version is existed in which gravity is the effect of the torsion not curvature. This parallel formulation of gravity in terms of the torsion called 
Teleparallel (TP) Theory. It's clear for us that  the equations of the motion (EOM) in TP is equivalent to the GR. This historically old equivalence  is called as 
teleparallel theory equivalent  with GR (TEGR) (for a review see\cite{TEGR}):
\begin{widetext}
\begin{eqnarray}&&TP\equiv\int T ed^4x=\int (2\nabla_{\mu}T^{\mu}d^4x-e\mathcal{R})\\&&\nonumber
=2\int_{\partial\mathcal{M}}T^{\mu}n_{\mu}d^3x-\int \mathcal{R}ed^4x\\&&\nonumber	\cong \int \mathcal{R}\sqrt{-g}d^4x\equiv GR\label{equiv}.
\end{eqnarray}
\end{widetext}
(here $e=det(e_{a}^{\mu})=\sqrt{-g}$). Note that starting from a generic extension of TP action, we can not obtain the same equivalent form with $f(R)$, i.e.:
\begin{widetext}
\begin{eqnarray}&&\int f(T) ed^4x\neq \int f(\mathcal{R})\sqrt{-g}d^4x.
\end{eqnarray}
\end{widetext}
where $\mathcal{R}$ denotes Ricci scalar is constracted from LC connections. 
The main reason is that despite the equivalency of TEGR given in (\ref{equiv}) , there is no bondary term $\int_{\partial\mathcal{M}}A^{\mu}n_{\mu}d^3x$. \par
 TP is not invariant under conformal transformations as well as Lorentz symmetry, although recently 
 a version has been proposed to preserve conformal symmetry\cite{maluf}(with a diffrent definition for torsion scalar $T$) , but the case of Lorentz symmetry is still an open problem in physics. The reason is the TP and its extensions all are frame dependent. To find a suitable frame, there is no manifest or program. If  one chooses wrong tetrads (frames) the results may be drastically changed. Such inequivalence frames are totally unfavoured. 
\par
A parallel extension of TP,  a new scenario proposed as gravity in the form of  $f(T)$, where here $T$ is torsion scalar is still frame dependent and consequently it violates the  Lorentz symmetry. The question will be as the following: is it possible to construct a Lorentz invariant theory of gravity in Weitzenb\"{o}ck spacetime?. Specially since $f(R)$ gravity is Lorentz invariant, is it possible to introduce a type of modified gravity in Weitzenb\"{o}ck spacetime like $f(T,X,...)$ which it respects to Lorentz symmetry?. We'll show that the unique type of such models will be a type of $f(\tilde{R})$ gravity in which $\tilde{R}$ is computed using the Weitzenb\"{o}ck connections in Weitzenb\"{o}ck spacetime. No such extension exist in which it includes $T,X,...$, where by $X$ we mean any additional field like $\nabla_{\mu}T^{\mu}$ as it was proposed in \cite{Bahamonde:2015zma}.
\par

%%%%%%%%%%%%%%%%%%%
\section{$f(T)$ gravity}
%%%%%%%%%%%%%%
Since we're interesting to study the Lorentz invariant extension of $f(R)$ in Weitzenb\"{o}ck spacetime, firstly we should quickly review the basic fundations of TP and its extension, $f(T)$. 
Let us to start by a gauge theory of gravity with  the physical Riemannian metric of theory in basis of tetrads one-forms $\theta^i$:
\begin{eqnarray}
ds^2=g_{\mu\nu}dx^\mu dx^\nu=\eta_{ij}\theta^i\theta^j\,,
\end{eqnarray}
The coordinate form (not unique) of this basis $\theta^i$ is as follows: 
\begin{eqnarray}
\theta^\mu=e_{i}^{\;\;\mu}\theta^{i}; \,\quad \theta^{i}=e^{i}_{\;\;\mu}dx^{\mu}.
\end{eqnarray}
Here the flat Minkowski metric is defined by $\eta_{ij}=diag(1,-1,-1,-1)$. For future purposes we define  $\{e^{i}_{\;\mu}\}$ by:
\begin{eqnarray}
e^{\;\;\mu}_{i}e^{i}_{\;\;\nu}=\delta^{\mu}_{\nu},\quad e^{\;\;i}_{\mu}e^{\mu}_{\;\;j}=\delta^{i}_{j}.
\end{eqnarray}
 In Weitzenb\"{o}ck spacetime the Weitzenb\"{o}ck connection is defined in a similar manner as the Levi-Civita connection:
\begin{eqnarray}
\Gamma^{\lambda}_{\mu\nu}=e^{\;\;\lambda}_{i}\partial_{\mu}e^{i}_{\;\;\nu}=-e^{i}_{\;\;\mu}\partial_\nu e_{i}^{\;\;\lambda}.
\end{eqnarray}
An object which will be analogoue to the  Ricci tensor, Riemann and others in Weitzenb\"{o}ck spacetime geometry are defined through the following lines:
\begin{eqnarray}
T^{\lambda}_{\;\;\;\mu\nu}= \Gamma^{\lambda}_{\mu\nu}-\Gamma^{\lambda}_{\nu\mu},
\end{eqnarray}
The one is called as the contorsion tensor :
\begin{eqnarray}
K^{\mu\nu}_{\;\;\;\;\lambda}=-\frac{1}{2}\left(T^{\mu\nu}_{\;\;\;\lambda}-T^{\nu\mu}_{\;\;\;\;\lambda}+T^{\;\;\;\nu\mu}_{\lambda}\right)\,\,\label{K}.
\end{eqnarray}
And the totally Lorentz violating tensor (frame dependent) $S_{\lambda}^{\;\;\mu\nu}$ :
\begin{eqnarray}
S_{\lambda}^{\;\;\mu\nu}=\frac{1}{2}\left(K^{\mu\nu}_{\;\;\;\;\lambda}+\delta^{\mu}_{\lambda}T^{\alpha\nu}_{\;\;\;\;\alpha}-\delta^{\nu}_{\lambda}T^{\alpha\mu}_{\;\;\;\;\alpha}\right)\,\,.
\end{eqnarray}
Thus, the  torsion scalar is constructed using these quantities:
\begin{eqnarray}
T=T^{\lambda}_{\;\;\;\mu\nu}S^{\;\;\;\mu\nu}_{\lambda}\,.
\end{eqnarray}
The general gravitational action for $f (T) $ in analogous to the $f (R) $ gravity is written as the following:
\begin{eqnarray}
 S= \int e \left[\frac{f(T)}{2\kappa^2} +\mathcal{L}_{m} \right]d^{4}x   \label{f(T)}\,,
\end{eqnarray}
Here as GR, $\kappa^{2} = 8 \pi G $.\par
One can derive the full system of the equation of the motion (EOM) using the  variation of the action (\ref{f(T)}) with respect to the basis tetrads:
\begin{widetext}
\begin{eqnarray}
S^{\;\;\; \nu \rho}_{\mu} \partial_{\rho} T f_{TT} + 
[e^{-1} e^{i}_{\;\; \mu}\partial_{\rho}(e e^{\;\; \mu}_{i}S^{\;\;\; \nu\lambda}_{\alpha} )
+T^{\alpha}_{\;\;\; \lambda \mu}   S^{\;\;\; \nu \lambda}_{\alpha} ]f_{T}+
\frac{1}{4}\delta^{\nu}_{\mu}f=\frac{\kappa^{2}}{2} \mathcal{T}^{\nu}_{\mu}  \label{eq10}\,,
\end{eqnarray}
\end{widetext}
As usual $\mathcal{T}^{\nu}_{\mu}$ is the energy 
momentum tensor of the matter Lagrangian $\mathcal{L}_{m} $, $f_{T} = df(T)/dT$ and 
$f_{TT}  = d^{2}f(T)/dT^{2}$. Due to the presence of the term $ S^{\;\;\; \nu \lambda}_{\alpha}$, $f(T)$ gravity is not Lorentz invariant. The results must be interpreted very carefully in respect to the frame (physical or unphysical) which are used. For this reason another type of modified gravity is proposed as an attempt to make it invariant \cite{Bahamonde:2015zma}. We'll review it in the forthcoming section.
\par

%%%%%%%%%%%%%%%%%%%%%%%%%%%%%%%%%%%%%%%%%%%%%%%%%%%%%%%%%%%%%%%%%%%%%%%%%%%%%
\section{ $f(T,\mathcal{B})$ Gravity }
%%%%%%%%%%%%%%%%%%%%%%%%%%%%%%%%%%%%%%%%%%%%%%%%%%%%%%%%%%%%%%%%%%%%%%%%%%%%%
In Weitzenb\"{o}ck spacetime , a simple computation showed that torsion scalar $T$, Ricci scalar $\mathcal{R}$ and tetrad fields gradient $\nabla_{\mu}T^{\mu}$ are related by the following algrbraic expression:
\begin{eqnarray}
&&\mathcal{R}+T=2\nabla_{\mu}T^{\mu}.
\end{eqnarray}
here Ricci scalar in Weitzenb\"{o}ck spacetime is defined by the Weitzenb\"{o}ck -Riemann curvature tensor:
\begin{eqnarray}
&&\mathcal{R}_{\rho\sigma\mu\nu}=R_{\rho\sigma\mu\nu}+\nabla_{\mu}
K_{\rho\sigma\nu}-\nabla_{\nu}K_{\rho\sigma\mu}+K_{\rho\lambda\mu}K^{\lambda}
_{\sigma\nu}-K_{\rho\lambda\nu}K^{\lambda}_{\sigma\mu}
,
\end{eqnarray}
here $K_{\rho\sigma\nu}$ is defined by (\ref{K}).
It will be amazing if we can construct a type of modified gravity like  $f(T,\mathcal{B})$, in  which $B=2\nabla_{\mu}T^{\mu}$. We would like to have a Lorentz invariant extension of $f(T)$ gravity. Such extension is reported \cite{Bahamonde:2015zma}. The gravitational action proposed by the following:

\begin{align}
  S_{\rm T \mathcal{B}} = \int 
  \left[ 
    \frac{1}{\kappa}f(T,\mathcal{B}) + L_{\rm m}
  \right] e\, d^4x \,,\label{f(T,B)}
\end{align}
where $f$ is an arbitrary  function of $T,\mathcal{B}$, and $L_{\rm m}$ is a matter Lagrangian. The corresponding EOM is obtained by the following 
variations of the action (\ref{f(T,B)}) with respect to the  $e_{a}^{\mu}$:
\begin{align}
  \delta S_{\rm T\mathcal{B}} = \int 
  \left[ 
    \frac{1}{\kappa}
    \Big(
    f(T,\mathcal{B})\delta e + e f_{\mathcal{B}}(T,\mathcal{\mathcal{B}}) 
\delta \mathcal{B} + e f_{T}(T,\mathcal{B}) \delta T
    \Big) 
    + \delta(e L_{\rm m})
  \right] 
  \, d^4x \,,
  \label{action2}
\end{align}
here 
\begin{align}
  ef_{\mathcal{B}}(T,\mathcal{B})\delta \mathcal{B} & = \Big[2eE_{a}^{\nu}\nabla^{\lambda}\nabla_{\mu}f_{\mathcal{B}}
-2eE_{a}^{\lambda}\Box f_{\mathcal{B}}-\mathcal{B}ef_{\mathcal{\mathcal{B}}}E_{a}^{\lambda}
-4e(\partial_{\mu}f_{\mathcal{B}})S_{a}\,^{\mu\lambda}\Big]\delta e_{\lambda}^{a} \,,
  \label{deltaB}\\
  ef_{T}(T,\mathcal{B})\delta T & = \Big[-4e(\partial_{\mu}f_{T})S_{a}\,^{\mu\lambda}-4\partial_{\mu}(e S_{a}\,^{\mu\lambda})f_{T}+4ef_{T}T^{\sigma}\,_{\mu a}S_{\sigma}\,^{\lambda\mu}\Big]\delta e^{a}_{\lambda} \,,
  \label{deltaT}\\
  f(T,\mathcal{B})\delta e & = e f(T,\mathcal{B})E_{a}^{\lambda} \delta e^{a}_{\lambda} 
  \label{deltae1} \,.
\end{align}
Plugging all equations, we find that the EOMs are given by the following "frame dependent" form \cite{Bahamonde:2015zma}:
\begin{multline}
  2eE_{a}^{\lambda}\Box f_{\mathcal{B}}-2eE_{a}^{\sigma}\nabla^{\lambda}
\nabla_{\sigma}f_{\mathcal{B}}+
  e\mathcal{B}f_{\mathcal{B}}E_{a}^{\lambda} + 4e\Big[(\partial_{\mu}f_{\mathcal{B}})+(\partial_{\mu}f_{T})\Big]S_{a}{}^{\mu\lambda} 
  \\
  +4\partial_{\mu}(e S_{a}{}^{\mu\lambda})f_{T}-4ef_{T}T^{\sigma}{}_{\mu a}S_{\sigma}{}^{\lambda\mu}-
  e f E_{a}^{\lambda} = 16\pi e \mathcal{T}_{a}^{\lambda}.
\end{multline}
By simple contracting  the above EOM with $e^{a}_{\nu}$ we obtain EOM in coordinates form:
\begin{multline}
  2e\delta_{\nu}^{\lambda}\Box f_{\mathcal{B}}-2e\nabla^{\lambda}\nabla_{\nu}f_{\mathcal{B}}+
  e \mathcal{B} f_{\mathcal{B}}\delta_{\nu}^{\lambda} + 
  4e\Big[(\partial_{\mu}f_{\mathcal{B}})
+(\partial_{\mu}f_{T})\Big]S_{\nu}{}^{\mu\lambda}
  \\
  +4e^{a}_{\nu}\partial_{\mu}(e S_{a}{}^{\mu\lambda})f_{T} - 
  4 e f_{T}T^{\sigma}{}_{\mu \nu}S_{\sigma}{}^{\lambda\mu} - 
  e f \delta_{\nu}^{\lambda} = 16\pi e\mathcal{T}_{\nu}^{\lambda} \,.
  \label{fieldeq}
\end{multline}
here $\mathcal{T}_{\nu}^{\lambda}=e^{a}_{\nu}\mathcal{T}_{a}^{\lambda}$ is defined as  the effective energy momentum tensor.  To have a better comparison with $f(R)$ gravity, 
 we can write EOM in the following equivalent but different form:
\begin{multline}
  H_{\mu\nu} := -f_{T}G_{\mu\nu}+g_{\mu\nu}\Box f_{\mathcal{B}}-\nabla_{\mu}\nabla_{\nu}f_{\mathcal{B}} +
  \frac{1}{2}(\mathcal{B}f_{\mathcal{B}}+Tf_{T}-f)g_{\mu\nu}
  \\
  +2\Big[(f_{\mathcal{B}\mathcal{B}}+f_{\mathcal{B}T})(\nabla_{\lambda}\mathcal{B})+(f_{TT}+f_{\mathcal{B}T})(\nabla_{\lambda}T)\Big]S_{\nu}{}^{\lambda}{}_{\mu}
  = 8\pi T_{\mu\nu} \,.
  \label{eq9}
\end{multline}

%%%%%%%%%%%%%%%%%%%%%%%%
\section{Lorentz invariant in $f(T,\mathcal{B})$}
%%%%%%%%%%%%%%%%%%%%%%%
Let us  firstly to study Lorentz symmetry in $f(T)$ gravity (\ref{f(T)}). Generally speaking, the torsion scalar $T$, depends only on the first derivatives of tetrads $\partial_{\nu}e_{a}^{\mu}$. By a different set of the tetrads, we obtain different set of the EOM. So, different local frames gives us different physics. It means $f(T)$ doesn't respect Lorentz symmetry \cite{Li:2010cg},\cite{Maluf:2013gaa}. Furthermore in the action of $f(T)$ gravity (\ref{f(T)})  we observe that 
 $f(T)$ doesn't differ from the $f(R)$ (\ref{f(R)}), consequently these theories are no longer equivalent like TEGR. The question is about the (\ref{f(T,B)}). Having look on the EOM given in (\ref{eq9}), the Lorentz violation term is $S_{\nu}{}^{\lambda}{}_{\mu}$. To remove it, one can set to zero the coefficent of this tensor term:
\begin{eqnarray}
&&(f_{\mathcal{B}\mathcal{B}}+f_{\mathcal{B}T})(\nabla_{\lambda}\mathcal{B})+(f_{TT}+f_{\mathcal{B}T})(\nabla_{\lambda}T)=0.
\end{eqnarray}
Both terms $\mathcal{B},T$ are frame dependent. So, we should omit them in a fully Lorentz invariant theory (\ref{f(T,B)}). It restricts the generic form of the model, since we must find a unique solution to the following pair of partial differential equations:
\begin{eqnarray}
&&f_{\mathcal{B}\mathcal{B}}+f_{\mathcal{B}T}=0,\ \ f_{TT}+f_{\mathcal{B}T}=0.
\end{eqnarray}
If we subtract this equation, we find a two dimensional wave equation with the following exact , "traveling" solution:
\begin{eqnarray}
&&f(T,\mathcal{B})=h(T+\mathcal{B})+k(T-\mathcal{B}).
\end{eqnarray}
 If we satisfy the first and second equations $f_{TT}=0$, by this form of f, we finally obtain:
\begin{eqnarray}
&&f(T,\mathcal{R})=c_1T+c_2+l(\mathcal{R}).
\end{eqnarray}
Now consider a similar equivalency as (\ref{equiv}):
\begin{widetext}
\begin{eqnarray}&&S_{f(T,\mathcal{B})}=\int (c_1T+c_2+l(\mathcal{R})) ed^4x\cong c_2\mathcal{V}_4-\int \mathcal{R}\sqrt{-g}d^4x+\int l(\mathcal{R}) \sqrt{-g}d^4x\\&&\nonumber \cong \int F(\mathcal{R}) \sqrt{-g}d^4x=S_{F(\mathcal{R})}
\label{equiv2}.
\end{eqnarray}
\end{widetext}
Here $\mathcal{V}_4\to\infty$ is the four volume of the entirely spacetime and we use (\ref{equiv}).
 We already showed that an $f $ in the  form given by Lorentz symmetry, simply reduces to the $f(\mathcal{R})$ field equations, which are manifestly Lorentz invariant. Therefore, the TP equivalent of $f(\mathcal{R}) $ gravity is the unique  Lorentz invariant theory of gravity constructed in Weitzenb\"{o}ck spacetime :
\begin{widetext}
\begin{eqnarray}&&S_{f(T,\mathcal{B})}+\text{Lorentz\ \ symmetry}=S_{F(\mathcal{R})}.
\end{eqnarray}
\end{widetext}
or equivalently we state that:
\begin{widetext}
\begin{eqnarray}&&\text{Modified\ \  gravity}+\text{Lorentz\ \ symmetry in \text{Weitzenb\"{o}ck\ \ spacetime }}=F(\mathcal{R}) .
\end{eqnarray}
\end{widetext}

%%%%%%%%%%%%%%%%%%%%%%%%%%%
 \section{Conclusion}
%%%%%%%%%%%%%%%%%%%%%%%%%%%%
In summary we study Lorentz symmetry is Weitzenb\"{o}ck spacetime. Because TP and its generalized form $f(T)$, don't respect Lorentz symmetry, we revisit a new type of modified gravity in Weitzenb\"{o}ck spacetime in which both $T$ and $\nabla_{\mu}T^{\mu}$ are appeared in action. We review equations of the motion of this theory. To keep Lorentz symmetry we need to specify the model. What we obtained was the only possible Lorentz invariant form of modified gravity by these ingredients, is $f(\mathcal{R})$ gravity. Here Ricci scalar is constructed from Weitzenb\"{o}ck connections. There is no way to address Lorentz symmetry in a model with $T$ or any other extra field. We conclude that even if we add an extra field $\nabla_{\mu}T^{\mu}$ to the TP theory , the model will not be Lorentz invariant. 
We should quote a former paper to the  \citep{Bahamonde:2015zma} in which the authors demonstrated that there is an interesting work in literature in which the authors considered  a scenario of modified gravity in which  the invariants $R$ and $T$ are dealt under the same standard in FRW cosmology \cite{fRT}. But to derive the correct cosmological background equations they introduced two auxiliary function $u,v$. The values of these functions should be matched by a phenomenological argument. In this approach we keep the same inspiration as we are studying in our paper but we need to imtroduce phenomenologically desired functions. Although it deserves more studies and it improves $f(T)$ gravity from a general point of view, the solution to the Lorentz violation still is unsolved.
Its interesting basically to study the weak field of our model like the $f(T)$ gravity \cite{wave}. In $f(T)$ gravity based on this important paper, the authors showed  that gravitational wave modes in purely torsion based model,  $f(T)$ gravity are matched to those modes in  GR.  The authors performed perturbation analysis for a class of analytic function of $f(T)$. Consequently no extra mode(s) (ghost or physical) emerges in $f(T)$. From the application point of view, by a gravitational detector we can't distinguse GR from $f(T)$ gravity. In this new scenario which we considered in our letter, such perturbation analysis is deserved to be consider in details. But it'll be beyond the scope of this letter. As a qualify analysis we can quote that in $f(T,\mathcal{B})$ model, because a vector mode is included, so we guess that this mode can deform polarization and from the deflection angle we can find some infomation about a possible gravitational wave which came from torsion of the spacetime $T$. This important issue will be addressed in a new forthcoming paper.

%{\bf Acknowledgement}: 


\begin{thebibliography}{17}

\addcontentsline{toc}{chapter}{Bibliographie}

%%%%%%%%%%%%%%%%%%
%f(T,B)$
%\cite{Bahamonde:2015zma}
\bibitem{Bahamonde:2015zma}
  S.~Bahamonde, C.~G.~Boehmer and M.~Wright,
  %``Modified teleparallel theories of gravity,''
  arXiv:1508.05120 [gr-qc].

%%%%%%%%%%%%%%%%
%Observational
%%%%%%%%%%%%%%%%%%
\bibitem{data}
S. Perlmutter et al. [SNCP Collaboration], Astrophys. J. {\bf 517}, 565 (1999); A. G. Riess et al.[SNST Collaboration], Astron. J. {\bf 116}, 1009 (1998);
D. N. Spergel et al. [WMAP Collaboration], Astrophys. J. Suppl. {\bf 148}, 175 (2003); ibid. 170,
377 (2007); E. Komatsu et al. [WMAP Collaboration], ibid. {\bf 180}, 330 (2009);
E. Komatsu et al. [WMAP Collaboration], Astrophys. J. Suppl. {\bf 192}, 18 (2011);
M. Tegmark et al., Phys. Rev. D {\bf 69}, 103501 (2004); U. Seljak et al. [SDSS Collaboration],
Phys. Rev. D {\bf 71}, 103515 (2005);
D. J. Eisenstein et al., Astrophys. J. {\bf 633}, 560 (2005);
B. Jain and A. Taylor, Phys. Rev. Lett. {\bf 91}, 141302 (2003).
%%%%%%%%%%%%%%%%%%%%%%%
%Modified gravity review
%%%%%%%%%%%%%%%%%%%%%%%
\bibitem{review}
S. Capozziello and M. De Laurentis, Phys. Rept. 509,
167 (2011); 
E. J. Copeland, M. Sami and S. Tsujikawa, Int. J. Mod. Phys. D 15, 1753 (2006);S.~'i.~Nojiri and S.~D.~Odintsov,
  %``Introduction to modified gravity and gravitational alternative for dark energy,''
  eConf C {\bf 0602061}, 06 (2006)
  [Int.\ J.\ Geom.\ Meth.\ Mod.\ Phys.\  {\bf 4}, 115 (2007)]
  [hep-th/0601213].
%%%%%%%%%%%%%%%%%%
%f(R)
%%%%%%%%%%%%%%%%%%%
\bibitem{Buchdahl:1983zz} 
  H.~A.~Buchdahl,
  %``Non-linear Lagrangians and cosmological theory,''
  Mon.\ Not.\ Roy.\ Astron.\ Soc.\  {\bf 150}, 1 (1970).

\bibitem{f(R)}
A.~De Felice and S.~Tsujikawa,
  %``f(R) theories,''
  Living Rev.\ Rel.\  {\bf 13} (2010) 3
  [arXiv:1002.4928 [gr-qc]];
S.~'i.~Nojiri and S.~D.~Odintsov,
  %``Unified cosmic history in modified gravity: from F(R) theory to Lorentz non-invariant models,''
  Phys.\ Rept.\  {\bf 505} (2011) 59
  [arXiv:1011.0544 [gr-qc]].


%%%%%%%%%%%%%%%%%%%%%%
%Gauss-Bonnet gravity
%%%%%%%%%%%%%%%%%%%%%%%%
\bibitem{GB}
G. Cognola, E. Elizade, S. Nojiri, S. D. Odintsov and S. Zerbini, 
%Dark energy in modified GB gravity: late-time acceleration and the hierarchy problem. 
Phys. Rev. D \textbf{73} 084007 (2006) [arxiv:hep-th/0601008];
  E.~Elizalde, R.~Myrzakulov, V.~V.~Obukhov and D.~S\'aez-G\'omez,
  %``LambdaCDM epoch reconstruction from F(R,G) and modified Gauss-Bonnet gravities,''
  Class. Quant. Grav. \textbf{27} (2010) 095007
  [arXiv:1001.3636 [gr-qc]];
A.~De Felice, J.~-M.~Gerard and T.~Suyama,
  %``Cosmological perturbation in f(R,G) theories with a perfect fluid,''
Phys.\ Rev.\ D {\bf 82}, 063526 (2010)
  % [arXiv:1005.1958 [astro-ph.CO]].
%%%%%%%%%%%%%%%%%%%%%%%%
%%%%%%%%%%%%%%%%%%%%%%
\bibitem{Nojiri:2005jg} 
  S.~'i.~Nojiri and S.~D.~Odintsov,
  %``Modified Gauss-Bonnet theory as gravitational alternative for dark energy,''
  Phys.\ Lett.\ B {\bf 631}, 1 (2005)
  [hep-th/0508049].
\bibitem{Cognola:2006eg} 
  G.~Cognola, E.~Elizalde, S.~'i.~Nojiri, S.~D.~Odintsov and S.~Zerbini,
  %``Dark energy in modified Gauss-Bonnet gravity: Late-time acceleration and the hierarchy problem,''
  Phys.\ Rev.\ D {\bf 73}, 084007 (2006)
  [hep-th/0601008].





%TEGR+f(T)
%%%%%%%%%%%%%%%%%%%%%%%%
\bibitem{TEGR}
F. W. Hehl, J. D. McCrea, E. W. Mielke and Y. Ne’eman, Phys. Rep. 258, 1 (1995).
\bibitem{maluf}
J. W. Maluf and F. F. Faria, Annalen Phys. 524, 366 (2012) [arXiv:1203.0040 [gr-qc]]..

%%%%%%%%%%%%%%%%%%%
%f(T)

%\cite{Bengochea:2008gz}
\bibitem{Bengochea:2008gz} 
  G.~R.~Bengochea and R.~Ferraro,
  %``Dark torsion as the cosmic speed-up,''
  Phys.\ Rev.\ D {\bf 79}, 124019 (2009)
  [arXiv:0812.1205 [astro-ph]].

%\cite{Linder:2010py}
\bibitem{Linder:2010py} 
  E.~V.~Linder,
  %``Einstein's Other Gravity and the Acceleration of the Universe,''
  Phys.\ Rev.\ D {\bf 81}, 127301 (2010)
  [Phys.\ Rev.\ D {\bf 82}, 109902 (2010)]
  [arXiv:1005.3039 [astro-ph.CO]].

 %%%%%%%%%%%%%%%%%%%%%%%%
%f(T,TAU)
%%%%%%%%%%%%%%%%%%%%%
%\cite{Harko:2014aja}
\bibitem{Harko:2014aja}
  T.~Harko, F.~S.~N.~Lobo, G.~Otalora and E.~N.~Saridakis,
  %``$f(T,\mathcal{T})$ gravity and cosmology,''
  arXiv:1405.0519 [gr-qc].

%%%%%%%%%%%%%%%%%%
%f(T,T_G)

%\cite{Kofinas:2014aka}
\bibitem{Kofinas:2014aka} 
  G.~Kofinas, G.~Leon and E.~N.~Saridakis,
  %``Dynamical behavior in $f(T,T_G)$ cosmology,''
  Class.\ Quant.\ Grav.\  {\bf 31}, 175011 (2014)
  [arXiv:1404.7100 [gr-qc]].






\bibitem{Li:2010cg} 
  B.~Li, T.~P.~Sotiriou and J.~D.~Barrow,
  %``$f(T)$ gravity and local Lorentz invariance,''
  Phys.\ Rev.\ D {\bf 83}, 064035 (2011)
  [arXiv:1010.1041 [gr-qc]].


%\cite{Maluf:2013gaa}
\bibitem{Maluf:2013gaa} 
  J.~W.~Maluf,
  %``The teleparallel equivalent of general relativity,''
  Annalen Phys.\  {\bf 525}, 339 (2013)
  [arXiv:1303.3897 [gr-qc]].

\bibitem{fRT}
Capozziello .S. and  De Laurentis. M., Myrzakulov. R.~,
  arXiv:1412.1471 [gr-qc]

\bibitem{wave}
  Bamba. K.~, Capozziello. S.~,  De Laurentis. M, Nojiri .S.~and Saez-Gomez. D.~,
  Phys.\ Lett.\ B {\bf 727} (2013) 194
\end{thebibliography}
\end{document}